\documentstyle[epsf]{article} 

\begin{document}            

\title{Spin-Peierls instability 
       in a quantum spin chain
       \protect\\
       with Dzyaloshinskii-Moriya interaction}

\author{
O. Derzhko$^{a}
\footnote{On leave of absence from
Institute for Condensed Matter Physics,
1 Svientsitskii Str., L'viv-11, 79011, Ukraine
and
Chair of Theoretical Physics, 
Ivan Franko National University in L'viv,
12 Drahomanov Str., L'viv-5, 79005, Ukraine}$, 
J. Richter$^{a}
\footnote{On leave of absence from
Institut f\"{u}r Theoretische Physik,
Universit\"{a}t Magdeburg,
P.O. Box 4120, D-39016 Magdeburg, Germany}$, 
and O. Zaburannyi$^{b}$\\
\small{$^{a}$Max-Planck-Institut f\"{u}r Physik komplexer Systeme,}\\
\small{N\"{o}thnitzer Str. 38, 01187 Dresden, Germany}\\
\small{$^{b}$Institute for Condensed Matter Physics,}\\
\small{1 Svientsitskii Str., L'viv-11, 79011, Ukraine}
}

\date{\today}

\maketitle                   

\begin{abstract}
We analysed the ground state energy 
of some dimerized spin-$\frac{1}{2}$ transverse $XX$ and Heisenberg chains
with Dzyaloshinskii-Moriya (DM) interaction
to study the influence of the latter interaction 
on the spin-Peierls instability.
We found that 
DM interaction may act 
either in favour of the dimerization or against it.
The actual result depends 
on the dependence of DM interaction 
on the distortion amplitude 
in comparison with such dependence for the isotropic exchange interaction.
\end{abstract}

\vspace{3mm}


\vspace{3mm}


\vspace{5mm}

\noindent
{\bf {Postal addresses:}}\\

\vspace{0mm}

\noindent
Dr. Oleg Derzhko (corresponding author)\\
Oles' Zaburannyi\\
Institute for Condensed Matter Physics\\
1 Svientsitskii Street, L'viv-11, 79011, Ukraine\\
tel/fax: (0322) 76 19 78\\
email: derzhko@icmp.lviv.ua\\

\vspace{0mm}

\noindent
Prof. Johannes Richter\\
Institut f\"{u}r Theoretische Physik,
Universit\"{a}t Magdeburg\\
P.O. Box 4120, D-39016 Magdeburg, Germany\\
tel: (0049) 391 671 8841\\
fax: (0049) 391 671 1217\\
email: Johannes.Richter@Physik.Uni-Magdeburg.DE

\clearpage

\renewcommand\baselinestretch{1.75}
\large\normalsize

The spin-Peierls instability is known 
as a magnetic analogue 
of the conventional Peierls instability in electron-phonon systems. 
A uniform quantum spin chain at low temperatures 
may become unstable towards dimerization 
owing to the interaction with lattice degrees of freedom.
This occurs 
because the dimerized lattice distortion 
lowers the magnetic energy by a greater amount 
than the increase in the elastic energy due to deformation.
Starting in the 70s with organic compounds 
exhibiting spin-Peierls transition,
the interest in the spin-Peierls instability of quantum spin chains 
was renewed with the discovery of the inorganic spin-Peierls compound 
CuGeO$_3$ in 1993 \cite{001,002}.
To model appropriately the spin degrees of freedom 
of the spin-Peierls compounds 
the pure Heisenberg chain 
as well as its modifications, 
which include frustration or interchain interaction, 
are considered.
As a rule, 
since those models 
represent quantum many body systems, 
only approximate results can be obtained.
However, some generic features of the spin-Peierls 
systems can be illustrated in a simplified but exactly solvable 
quantum spin model,  
namely, the transverse $XX$ chain
\cite{003,004,005,006}.

In the present paper 
we discuss the influence
of the Dzyaloshinskii-Moriya (DM) interaction \cite{007}
on the
spin-Peierls dimerization in the adiabatic limit.
The presence of DM interaction  
for CuGeO$_3$ was proposed
in Refs. \cite{008,009,010}
in order to explain the EPR and ESR experimental data.
The structure of DM interaction in the cuprates 
was examined in Refs. \cite{011,012}. 
The influence of this interaction on the ground state properties 
of the one-dimensional and
two-dimensional Heisenberg models 
was studied in Refs. \cite{013,014,015}.
Besides, 
the multisublattice transverse $XX$ chain with DM interaction 
was introduced in Ref. \cite{016},
however, the spin-Peierls instability was discussed 
only in one limiting case (see below) in the absence of an external field.
DM interaction was found  
to be present in a number of quasi-one-dimensional magnets 
(see, e.g., \cite{017,018} 
and also \cite{019} 
in which RbCoCl$_3$$\cdot$2H$_2$O is described as a pure DM chain)
and a study of one-dimensional DM Hamiltonians 
seems to be of great importance.

Several mechanisms which may destroy the dimerized phase 
in the Heisenberg chain 
were discussed in the literature, 
in particular, 
an external field or an Ising anisotropy.
$XX$ anisotropy also suppresses the dimerized phase 
although does not destroy it completely. 
Therefore, initially one may expect that the appearance of an anisotropy 
in form of the DM interaction 
could act against dimerization. 
Although we shall find that the Heisenberg chain 
with increasing of DM interaction 
may become similar to the $XX$ chain 
both enhancing or suppressing of the dimerized phase are possible 
depending on the details of the distortion dependence of DM interaction.

In our study we follow the idea of
Ref. \cite{003}
and compare the total ground state energy
of the dimerized and uniform chains
in the presence of DM interaction. 
First we demonstrate 
that in some cases the DM interaction 
can be eliminated 
by a spin coordinate transformation 
resulting in a model with
an anisotropic exchange interaction. 
This observation permits us 
to study rigorously the influence of DM interaction 
on the spin-Peierls instability
in the transverse $XX$ chain  
using the exact results 
for thermodynamic quantities of the regularly alternating  
transverse $XX$ chain 
obtained recently with the help of continued fractions \cite{020}.  
(The approach exploiting continued fractions 
in contrast to the approaches used in previous works
\cite{003,004,005,006,016}
allows one
to consider in a similar way not only the dimerized lattice but
also more complicated lattice distortions.)  
Further, 
we discuss the case of the Heisenberg chain 
with DM interaction 
using exact diagonalization of finite chains. 
The exact analytical findings for the $XX$ chain 
are helpful for the interpretation of
the finite-chain results for more realistic Heisenberg chain. 
In the present study we are interested 
in describing the generic features 
originated by DM interaction   
and therefore no compound specific parameters are considered. 

To begin with, 
we consider a nonuniform chain of $N\rightarrow\infty$ spins $\frac{1}{2}$ 
governed by the isotropic Heisenberg Hamiltonian 
with DM interaction
\begin{eqnarray}
\label{01}
H=\sum_n\Omega_ns^z_n
+\sum_n
\left(
J_n({\bf s}_n\cdot{\bf s}_{n+1})
+{\bf D}_n\cdot[{\bf s}_n\times{\bf s}_{n+1}]
\right).
\end{eqnarray}
Here $\Omega_n$ is an external field at site $n$,
and $J_n$ and ${\bf D}_n$ 
are the isotropic exchange interaction 
and the antisymmetric anisotropic exchange interaction 
or DM interaction 
between the neighbouring sites $n$ and $n+1$, respectively.

Extending the spin coordinate transformation used in Ref. \cite{021} 
for models with nonuniform $J_n$ and ${\bf{D}}_n$ 
first we show 
how the terms with cross product can be eliminated from Hamiltonian ({\ref{01}) 
in special cases.
If ${\bf D}_n$ has only one nonzero component $D^z_n$ 
one may perform local rotations about the $z$ axis 
introducing new spin operators 
${s^x_n}^{\prime}=s^x_n\cos{\phi_n}+s^y_n\sin{\phi_n}$,
${s^y_n}^{\prime}=-s^x_n\sin{\phi_n}+s^y_n\cos{\phi_n}$,
$\phi_n=\varphi_1+\ldots+\varphi_{n-1}$,
$\tan\varphi_m=\frac{D^z_m}{J_m}$ 
in terms of which (\ref{01}) becomes
\cite{022}
\begin{eqnarray}
\label{02}
H=\sum_n\Omega_n{s^z_n}^{\prime}
+\sum_n
\left(
\sqrt{J_n^2+{D^z_n}^2}
\left({s^x_n}^{\prime}{s^x_{n+1}}^{\prime}
+{s^y_n}^{\prime}{s^y_{n+1}}^{\prime}\right)
+J_n{s^z_n}^{\prime}{s^z_{n+1}}^{\prime}
\right).
\end{eqnarray}
Note, that such transformation can be also applied 
to the transverse $XX$ chain 
in which $z$ spin components do not interact.  
Assume further that 
${\bf D}_n$ has only one nonzero component $D^x_n$. 
Then the local rotations must be performed about the $x$ axis, 
i.e. 
${s^y_n}^{\prime}=s^y_n\cos{\phi_n}+s^z_n\sin{\phi_n}$,
${s^z_n}^{\prime}=-s^y_n\sin{\phi_n}+s^z_n\cos{\phi_n}$
with
$\tan\varphi_m=\frac{D^x_m}{J_m}$ 
resulting in
\begin{eqnarray}
\label{03}
H=
\sum_n
\Omega_n
\left(\sin\phi_n{s^y_n}^{\prime}+\cos\phi_n{s^z_n}^{\prime}\right)
+\sum_n
\left(
J_n{s^x_n}^{\prime}{s^x_{n+1}}^{\prime}
+\sqrt{J_n^2+{D^x_n}^2}
\left({s^y_n}^{\prime}{s^y_{n+1}}^{\prime}
+{s^z_n}^{\prime}{s^z_{n+1}}^{\prime}\right)
\right).
\end{eqnarray}
Acting similarly for ${\bf D}_n=(0,D_n^y,0)$
one finds that the transformed Hamiltonian is given by (\ref{03}) 
with the replacement 
$D^x_n\to D^y_n$,
${s^x}^{\prime}\to{s^y}^{\prime}$, 
${s^y}^{\prime}\to -{s^x}^{\prime}$.
In the case when ${\bf D}_n$ has more than one nonzero components 
the described elimination can be performed 
if the orientation of ${\bf D}_n$ 
(but not necessarily its value 
$D_n=\sqrt{{D^x_n}^2+{D^y_n}^2+{D^z_n}^2}\;$)
is site independent,
i.e. 
$\frac{D^x_n}{D_n}$, 
$\frac{D^y_n}{D_n}$, 
$\frac{D^z_n}{D_n}$
do not depend on $n$.
(Obviously, the particular site independent orientations of ${\bf D}_n$ 
reproduce the cases discussed above.)
Really, 
in such a case we start from the global transformation of coordinate system 
with the Eulerian angles $\phi$, $\theta$, $\psi$ putting 
$\tan\phi=\frac{D^y_n}{D^x_n}$,
$\theta=\frac{\pi}{2}$,
$\tan\psi=\frac{D^z_n}{\sqrt{{D^x_n}^2+{D^y_n}^2}}$
and then perform the mentioned above local rotations about the $x$ axis 
with $\tan\varphi_m=\frac{D_m}{J_m}$ 
finding as a result
\begin{eqnarray}
\label{04}
H=
\sum_n
\Omega_n
\left(
\frac{D^z_n}{D_n}{s^x_n}^{\prime}
+\frac{\sqrt{{D^x_n}^2+{D^y_n}^2}}{D_n}
\left(\cos\phi_n{s^y_n}^{\prime}
-\sin\phi_n{s^z_n}^{\prime}\right)
\right)
\nonumber\\
+\sum_n
\left(
J_n{s^x_n}^{\prime}{s^x_{n+1}}^{\prime}
+\sqrt{J_n^2+{D_n}^2}
\left({s^y_n}^{\prime}{s^y_{n+1}}^{\prime}
+{s^z_n}^{\prime}{s^z_{n+1}}^{\prime}\right)
\right).
\end{eqnarray}
It should be noted that the described transformations 
were applied to open chains.
For cyclic chains 
they yield the presented expressions (\ref{02}) -- (\ref{04})  
at least up to the boundary term.  
Evidently, the thermodynamic properties 
of the initial and transformed Hamiltonians are identical.
Due to these transformations (Eqs. (\ref{02}) -- (\ref{04})) 
one is able to exploit the broad knowledge on 
anisotropic Heisenberg chains and $XX$ chains.
In what follow we shall use the thermodynamic equivalence
of the initial and transformed Hamiltonians  
both in the analytical treatment and numerical computations.
 
We proceed considering 
the transverse $XX$ chain with DM interaction 
having only $z$ component 
\begin{eqnarray}
\label{05}
H=\sum_{n}\Omega_ns_n^z
+\sum_{n}J_n\left(s_n^xs_{n+1}^x+s_n^ys_{n+1}^y\right)
+\sum_{n}D_n^z\left(s_n^xs_{n+1}^y-s_n^ys_{n+1}^x\right).
\end{eqnarray}
For the Hamiltonian (\ref{05}) 
we are able to perform rigorous analytical calculations 
since 
after the Jordan-Wigner transformation 
it reduces to noninteracting spinless fermions.  
As it has been already mentioned  
thermodynamics of the model given by (\ref{05}) 
is the same as of the transverse $XX$ chain 
(without DM interaction) 
with exchange interaction $\sqrt{J_n^2+{D^z_n}^2}$. 
The thermodynamic properties 
of the latter model with regular alternation in bonds and fields 
having finite period $p$  
have been examined recently 
with the help of continued fractions \cite{020}. 
To study the spin-Peierls dimerization in the adiabatic  limit 
we need the ground state energy of a spin chain with period $p=2$,
i.e. with the sequence of parameters
$\Omega_1J_1D^z_1\Omega_2J_2D^z_2
\Omega_1J_1D^z_1\Omega_2J_2D^z_2\dots\;$. 
Moreover, we assume
the following reasonable dimerization ansatz 
$J_1=J(1+\delta)$,
$J_2=J(1-\delta)$,
$D_1^z=D^z(1+k\delta)$,
$D_2^z=D^z(1-k\delta)$,
where
$0\le\delta\le 1$
is the dimerization parameter.
It is argued  
that the directions of ${\bf{D}}$-vectors 
are not changed by the dimerization \cite{010,023}. 
From Ref. \cite{007} we know 
that the dependence on the intersite distance 
of the isotropic exchange interaction 
and DM interaction may be different.
This effect is described by the parameter $k$.
Putting $k=0$ one has a chain in which $D^z$ 
does not depend on the lattice distortion, whereas for $k=1$ the dependence 
of $D^z$ on the lattice distortion is as 
that for the isotropic exchange interaction $J$.
The latter case with $\Omega_n=0$ was considered in \cite{016}.
Besides, we bear in mind 
that, 
as a rule,
the value of DM interaction 
is significantly smaller than 
the value of isotropic exchange interaction 
\cite{007}. 

Further,
we consider the case of zero temperature 
and look for the total energy per site
${\cal{E}}(\delta)$
which consists of the magnetic part 
$e_0(\delta)$
and the elastic part
$\alpha\delta^2,\;\alpha>0$. 
From Ref. \cite{020} 
we know the exact expression for the magnetic ground state energy 
\begin{eqnarray}
\label{06}
e_0(\delta)
=-\frac{1}{\pi}{\sf{b}}_1
{\mbox{E}}\left(\psi,\frac{{\sf{b}}_1^2-{\sf{b}}_2^2}{{\sf{b}}_1^2}\right)
-\frac{1}{2}
\left\vert\Omega_1+\Omega_2\right\vert
\left(
\frac{1}{2}-\frac{\psi}{\pi}
\right),
\\
{\sf{b}}_{1,2}
=\frac{1}{2}
\sqrt{\left(\Omega_1-\Omega_2\right)^2
+\left(I_1\pm I_2\right)^2},
\nonumber\\
I_{1,2}
=\sqrt{J^2(1\pm\delta)^2+{D^z}^2(1\pm k\delta)^2},
\nonumber\\
\psi
=\left\{
\begin{array}{ll}
0, & {\mbox{if}} \;\;\; 
{\sf{b}}_1\le\frac{1}{2}\vert\Omega_1+\Omega_2\vert,\\
{\mbox{arcsin}}\sqrt{
\frac{{\sf{b}}_1^2-\frac{1}{4}\left(\Omega_1+\Omega_2\right)^2}
{{\sf{b}}_1^2-{\sf{b}}_2^2}}, &
{\mbox{if}} \;\;\; 
{\sf{b}}_2\le\frac{1}{2}\vert\Omega_1+\Omega_2\vert<{\sf{b}}_1,\\
\frac{\pi}{2}, &  {\mbox{if}} \;\;\;  
\frac{1}{2}\vert\Omega_1+\Omega_2\vert<{\sf{b}}_2,
\end{array}
\right.
\nonumber
\end{eqnarray}
where
${\mbox{E}}(\psi,a^2)\equiv
\int_0^{\psi}d\phi\sqrt{1-a^2\sin^2\phi}$
is the elliptic integral of the second kind. 
We also seek for a nonzero solution 
$\delta^{\star}\ne 0$
of the equation
$\frac{\partial{\cal{E}}(\delta)}{\partial\delta}=0$
that can be easily derived from (\ref{06}).
In what follows we consider the case of a uniform 
transverse field
$\Omega_1=\Omega_2=\Omega_0\ge 0$.
In the limit
$\delta\ll 1$
valid for hard lattices 
(having large values of $\alpha$ 
and corresponding to the experimental situation) 
one finds
${\sf{b}}_{1}=I$,
${\sf{b}}_{2}=I\aleph\delta$
with
$I=\sqrt{J^2+{D^z}^2}$
and
$\aleph=\frac{J^2+k{D^z}^2}
{J^2+{D^z}^2}$.
Instead of Eq. (\ref{06})
one then has
\begin{eqnarray}
\label{07}
e_0(\delta)=-\frac{I}{\pi}
{\mbox{E}}(\psi, 1-\aleph^2\delta^2)
-\Omega_0\left(\frac{1}{2}-\frac{\psi}{\pi}\right),
\\
\psi=
\left\{
\begin{array}{ll}
0, & 
{\mbox{if}}\;\;\;I<\Omega_0,\\
{\mbox{arcsin}}\sqrt{\frac{I^2-\Omega_0^2}
{I^2(1-\aleph^2\delta^2)}}, &
{\mbox{if}}\;\;\;I\aleph\delta\le\Omega_0<I,\\
\frac{\pi}{2}, &
{\mbox{if}}\;\;\;\Omega_0<I\aleph\delta,
\end{array}
\right.
\nonumber
\end{eqnarray}
whereas the equation for $\delta^{\star}$ reads
\begin{eqnarray}
\label{08}
\frac{2\pi\alpha}{I}
=\frac{\aleph^2}{1-\aleph^2{\delta^{\star}}^2}
\left(F(\psi,1-\aleph^2{\delta^{\star}}^2)
-E(\psi,1-\aleph^2{\delta^{\star}}^2)\right),
\end{eqnarray}
where 
${\mbox{F}}(\psi,a^2)\equiv\int_0^{\psi}\frac{d\phi}
{\sqrt{1-a^2\sin^2\phi}}$
is the elliptic integral of the first kind.

Consider at first the case $\Omega_0=0$.
After rescaling
$I\rightarrow J$,
$\frac{\alpha}{\aleph^2}\rightarrow\alpha$,
$\aleph\delta^{\star}\rightarrow\delta^{\star}$
one finds that 
Eq. (\ref{08})
is exactly the same as considered in Ref. \cite{003} and thus
$\delta^{\star}\sim \frac{1}{\aleph}
\exp
\left(
-\frac{2\pi\alpha}{I\aleph^2}
\right)$.
Thus 
for $k=1$ ($\aleph=1$)
nonzero $D^z$ leads to an increasing of
the dimerization parameter
$\delta^{\star}$, 
whereas for $k=0$ ($\aleph\le 1$)
nonzero $D^z$ leads to a decreasing of 
$\delta^{\star}$.
Let us pass to the case
$0<\Omega_0<I$.
Varying $\delta^{\star}$ in the r.h.s. of Eq. (\ref{08}) from 0 to 1
one calculates a lattice parameter $\alpha$
for which the taken value of $\delta^{\star}$ realizes an extremum of
${\cal{E}}(\delta)$ (\ref{07}).
One immediately observes that for
$\frac{\Omega_0}{I\aleph}\le\delta^{\star}$
the dependence
$\alpha$
versus
$\delta^{\star}$
remains as that in the absence of the field,
whereas for
$0\le\delta^{\star}<\frac{\Omega_0}{I\aleph}$
the calculated quantity
$\alpha$
starts to decrease.
From this one concludes that 
the field
$\frac{\Omega_0}{I}
=\exp\left(-\frac{2\pi\alpha}{I\aleph^2}\right)$
makes the dimerization
unstable against the uniform phase.
The latter relation tells us 
that nonzero $D^z$
increases the value of that field for $k=1$ and decreases it for $k=0$.
It is known
\cite{002}
that the increasing of
the external field leads to a transition from the dimerized phase
to the incommensurate phase rather than to the uniform phase.
However, the former phase cannot appear
within the frames of the adopted ansatz for the
lattice distortions
$\delta_1\delta_2\delta_1\delta_2\ldots\;$,
$\delta_1+\delta_2=0$.

After the discussion of the limit $\delta\ll 1$
we now present the results 
for arbitrary $0\le\delta\le 1$ 
based on (\ref{06}).
In Figs. 1, 2 we plot the changes of the total energy 
${\cal{E}}(\delta)-{\cal{E}}(0)$
(\ref{06}) vs $\delta$ 
and the nonzero solution $\delta^{\star}$ of equation
$\frac{\partial{\cal{E}}(\delta)}{\partial\delta}=0$ 
vs $\alpha$, 
respectively,  
for various strengths of 
DM interaction. 
These results 
confirm that
for $k=1$ DM interaction $D^z$ acts in favour of dimerization, 
whereas for $k=0$ against it.
To understand the validity of the data  
obtained by exact diagonalization of finite chains 
that will be used below 
for the Heisenberg chain 
we present in Fig. 1 
also the numerical results 
${\cal{E}}(\delta)-{\cal{E}}(0)$ vs $\delta$ 
for $N=24$ spins 
(open boundary conditions). 
Note, that although the finite chain results for $N=24$ still 
overestimate noticeably 
the value of $\delta^{\star}$ 
and the depth ${\cal{E}}(\delta^{\star})-{\cal{E}}(0)$,
however, 
and this is most important,
they reproduce qualitatively correctly
the influence of $D^z$ 
in both cases $k=1$ and $k=0$. 
The possible influences of $D^z$ are reproduced correctly 
even for shorter chains of $N=16,\;20$ spins.
Moreover, 
with increasing the chain length 
from $N=16$ to $24$
the numerical data 
approach the analytical ones
valid for $N\to\infty$. 
This expected tendency can be also seen 
by comparison curves 1 for $N=24$, $N=28$ 
(the computation in this case becomes already very time consuming) 
and $N\to\infty$ in Fig. 1.

Let us turn to the Heisenberg chain with DM interaction. 
Assume 
that the vectors ${\bf{D}}_n$ have the same orientation at all sites,
e.g. in the $z$ direction.
(The assumption 
${\bf{D}}_n=(0,0,D^z_n)$
does not lead to a loss of generality if $\Omega_0=0$.)
In such case 
the influence of DM interaction on thermodynamics 
follows from a study of thermodynamic properties 
of the Heisenberg chain 
with anisotropic exchange interaction ({\ref{02}). 
We immediately find 
the appearance of $XX$ anisotropy, 
since 
the interaction between $x$ and $y$ spin components becomes  
$\sqrt{J^2(1\pm\delta)^2+D^2(1\pm k\delta)^2}$ 
($\approx\sqrt{J^2+D^2}(1\pm\aleph\delta)$ for $\delta\ll 1$)
whereas between $z$ spin components it remains
$J(1\pm\delta)$.
Restricting ourselves to small $\delta$ 
we perform the redefinitions 
$\sqrt{J^2+D^2}=J^{\prime}$,
$\aleph\delta=\delta^{\prime}$,
$\frac{\alpha}{\aleph^2}=\alpha^{\prime}$. 
As a result we come to the anisotropic alternating Heisenberg chain 
with $xx$ and $yy$ interactions 
$J^{\prime}(1\pm\delta^{\prime})$ 
and $zz$ interaction 
$J^{\prime}\Delta(1\pm\frac{\delta^{\prime}}{\aleph})$, 
$\Delta=\frac{J}{\sqrt{J^2+D^2}}$ 
and the expression for the elastic energy per site  
$\alpha^{\prime}{\delta^{\prime}}^2$.
Any increase of $D$ results in increasing of intersite coupling $J^{\prime}$ 
and increasing of $XX$ anisotropy 
manifested by going of $\Delta$ from 1 ($D=0$) to 0 ($D\to\infty$).
Hence, 
the Heisenberg chain should start to
exhibit a behaviour inherent in the $XX$ chain 
while $D$ becomes large. 
Omitting the role of a change in the $zz$ interaction, 
which becomes less important while $D$ increases, 
and bearing in mind the corresponding analysis for the $XX$ chain 
based on (\ref{08}),    
we may expect 
that for $\Omega_0=0$
the dimerized phase will be enhanced 
for $k=1$ (since $\alpha^{\prime}=\alpha$)
and will be suppressed  
for $k=0$ (since $\alpha^{\prime}>\alpha$).
The numerical results for 
${\cal{E}}(\delta)-{\cal{E}}(0)$ vs $\delta$ 
for the Heisenberg chain of $N=24$ sites 
(open boundary conditions) 
shown in Fig. 3 confirm this expectation. 
As can be seen from the displayed plots 
the general tendency 
for the changes caused by DM interaction 
is the same for $XX$ and Heisenberg chains.

To conclude, we have examined the stability 
of some spin-$\frac{1}{2}$ transverse $XX$ and Heisenberg chains 
with respect to dimerization in the presence of
DM interaction
analysing the dependence 
of the ground state energy on dimerization parameter.
If the orientations of ${\bf{D}}_n$ are the same at all sites 
essential simplification occurs, 
i.e. the terms with cross product can be eliminated from the Hamiltonian 
resulting in the appearance of the anisotropy in the exchange interaction.
In addition, the external field becomes more complicated 
having three site dependent components (see (\ref{04})).  
The transformed Hamiltonian may be more transparent and convenient  
for further analytical or numerical treatment. 
For the transverse $XX$ chain we have found 
that DM interaction 
having only $z$ component 
may act both in favour of the dimerization or against it. 
The result of its influence depends on the dependence 
of DM interaction on the amplitude of lattice distortion
in comparison with a corresponding  dependence 
of the isotropic exchange interaction.
For the Heisenberg chain 
DM interaction having the same orientation at all sites
leads to $XX$ anisotropy 
(and extra nonuniform on-site fields in the presence of an external field).
For the Heisenberg chain without field 
we have observed qualitatively the same behaviour 
as in the $XX$ chain 
determined by the dependence of DM interaction on the lattice distortion. 

\vspace{5mm}

O.~D. and J.~R. acknowledge the kind hospitality 
of the Max Planck Institute for the Physics of Complex Systems, Dresden 
in the end of 1999
when a main part of the paper was done.
The present study was partly supported by the DFG 
(projects 436 UKR 17/20/98 and Ri 615/6-1).

\vspace{15mm}

FIGURE CAPTURES

\vspace{2mm}

FIGURE 1.
Dependence 
${\cal{E}}(\delta)-{\cal{E}}(0)$ (\ref{06}) vs $\delta$
for the transverse $XX$ chain with DM interaction. 
$J=2$, 
$\Omega_0=0$,
$\alpha=0.8$, 
$D^z=0$ (curve 1),
$D^z=0.4,\;0.8$
for $k=1$ 
(curves 2 and 3, respectively) 
and $k=0$ 
(curves 4 and 5, respectively).
Solid curves correspond to analytical calculations
($N\to\infty$),
whereas dotted ones to exact diagonalization results
($N=24$). 
The dashed curve corresponds to the numerical results for $N=28$.

\vspace{2mm}
 
FIGURE 2.
Dependence $\delta^{\star}$ vs $\alpha$ 
for the transverse $XX$ chain with DM interaction. 
$J=2$,
$\alpha=0.8$,
$\Omega_0=0$ (a),
$\Omega_0=0.2$ (b),
$D^z=0$ (curve 1),
$D^z=0.4,\;0.8$
for $k=1$ 
(curves 2 and 3, respectively) 
and $k=0$ 
(curves 4 and 5, respectively).

\vspace{2mm}
 
FIGURE 3.
Dependence 
${\cal{E}}(\delta)-{\cal{E}}(0)$ vs $\delta$
for the Heisenberg chain with DM interaction 
obtained by exact diagonalization of finite chains. 
$J=2$, 
$\Omega_0=0$,
$\alpha=0.8$, 
$D=0$ (curve 1),
$D=0.4,\;0.8$
for $k=1$ 
(curves 2 and 3, respectively) 
and $k=0$ 
(curves 4 and 5, respectively).
We also plotted the dependence 
${\cal{E}}(\delta)-{\cal{E}}(0)$ vs $\delta$
for the corresponding $XX$ chain with $D=0$ 
(i.e. curves 1 in Fig. 1) 
as it follows from 
analytical calculation for $N\to\infty$ (dashed curve)
and
numerical computation for $N=24$ (dotted curve).

\clearpage

\begin{figure}[t]
\epsfysize=100mm
\epsfclipon
\centerline{\epsffile{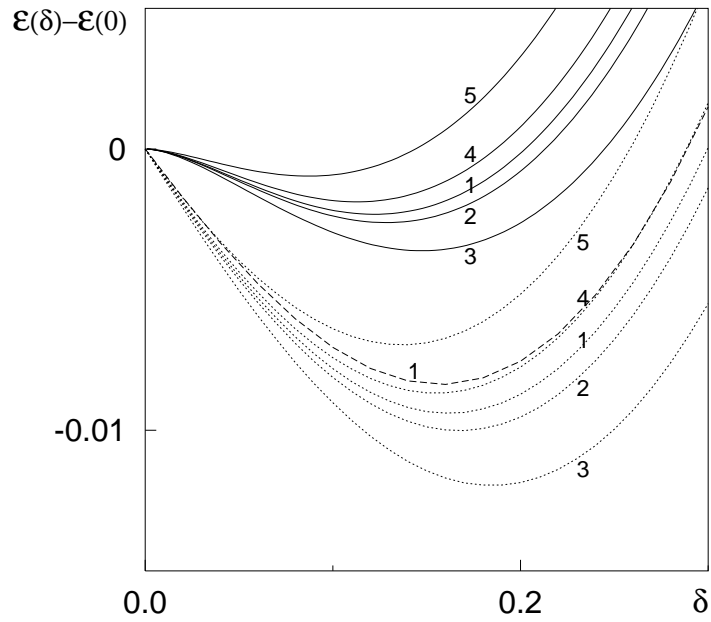}}
\caption[]
{\small
Dependence 
${\cal{E}}(\delta)-{\cal{E}}(0)$ (\ref{06}) vs $\delta$
for the transverse $XX$ chain with DM interaction. 
$J=2$, 
$\Omega_0=0$,
$\alpha=0.8$, 
$D^z=0$ (curve 1),
$D^z=0.4,\;0.8$
for $k=1$ 
(curves 2 and 3, respectively) 
and $k=0$ 
(curves 4 and 5, respectively).
Solid curves correspond to analytical calculations
($N\to\infty$),
whereas dotted ones to exact diagonalization results
($N=24$).
The dashed curve corresponds to the numerical results for $N=28$.}
\label{fig1}
\end{figure}

\clearpage

\begin{figure}[t]
\epsfysize=160mm
\epsfclipon
\centerline{\epsffile{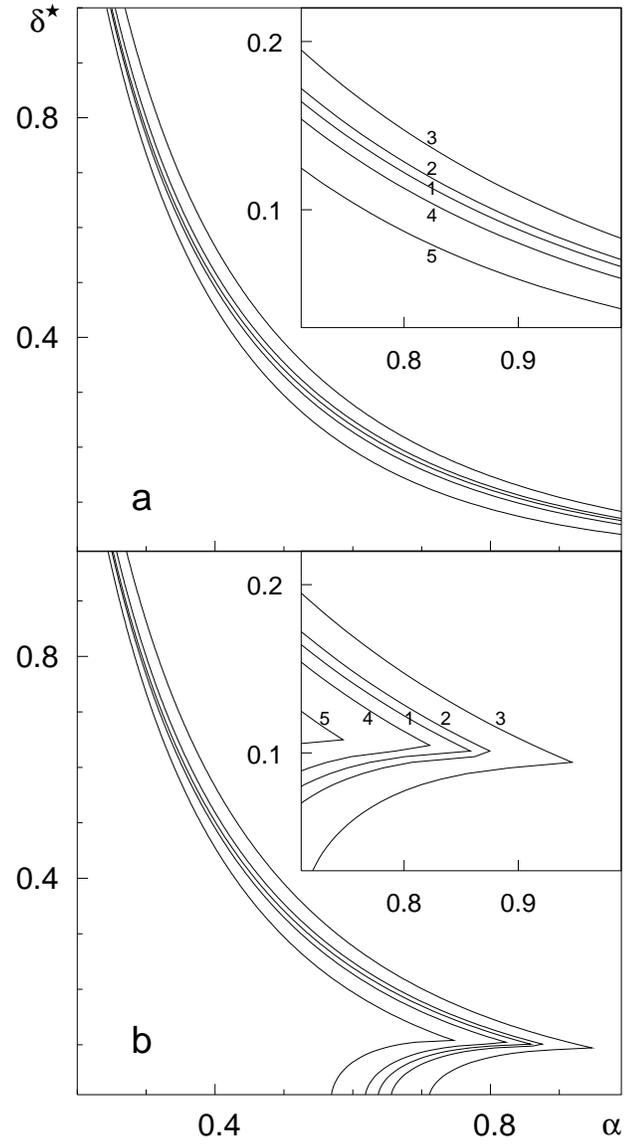}}
\caption[]
{\small
Dependence $\delta^{\star}$ vs $\alpha$ 
for the transverse $XX$ chain with DM interaction. 
$J=2$,
$\alpha=0.8$,
$\Omega_0=0$ (a),
$\Omega_0=0.2$ (b),
$D^z=0$ (curve 1),
$D^z=0.4,\;0.8$
for $k=1$ 
(curves 2 and 3, respectively) 
and $k=0$ 
(curves 4 and 5, respectively).}
\label{fig2}
\end{figure}

\clearpage

\begin{figure}[t]
\epsfysize=100mm
\epsfclipon
\centerline{\epsffile{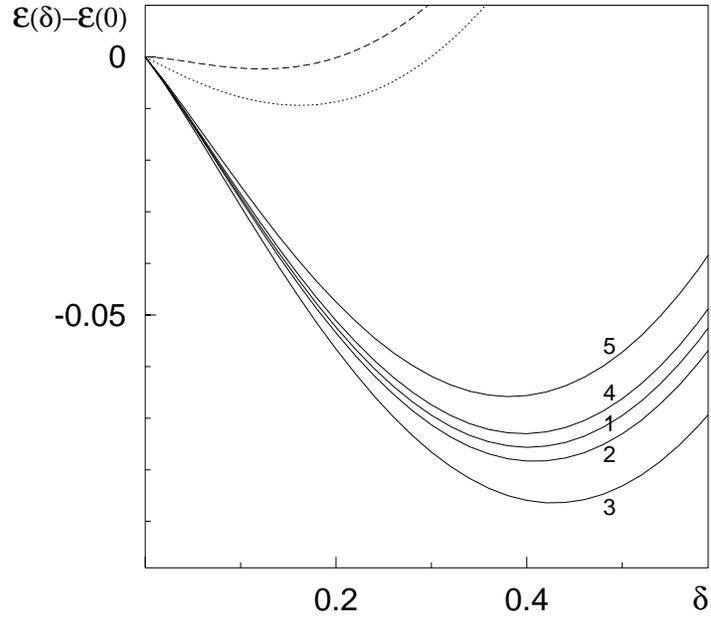}}
\caption[]
{\small
Dependence 
${\cal{E}}(\delta)-{\cal{E}}(0)$ vs $\delta$
for the Heisenberg chain with DM interaction 
obtained by exact diagonalization of finite chains. 
$J=2$, 
$\Omega_0=0$,
$\alpha=0.8$, 
$D=0$ (curve 1),
$D=0.4,\;0.8$
for $k=1$ 
(curves 2 and 3, respectively) 
and $k=0$ 
(curves 4 and 5, respectively).
We also plotted the dependence 
${\cal{E}}(\delta)-{\cal{E}}(0)$ vs $\delta$
for the corresponding $XX$ chain with $D=0$ 
(i.e. curves 1 in Fig. 1) 
as it follows from 
analytical calculation for $N\to\infty$ (dashed curve)
and
numerical computation for $N=24$ (dotted curve).}
\label{fig3}
\end{figure}

\end{document}